\documentclass[twocolumn,superscriptaddress,showpacs]{revtex4}
\usepackage{graphicx}
\usepackage{amsmath}

\begin{document}
\title{Restoring site percolation on a damaged square lattice}

\author{Serge Galam}
\email{galam@ccr.jussieu.fr}
\affiliation{
Universit\'e Pierre et Marie Curie et CNRS,
Laboratoire des Milieux D\'esordonn\'es et H\'et\'erog\`enes\\
Case 86, 4 place Jussieu, F-75252 Paris Cedex 05, France
}

\author{Krzysztof Malarz}
\homepage{http://home.agh.edu.pl/malarz/}

\affiliation{
Universit\'e Pierre et Marie Curie et CNRS,
Laboratoire des Milieux D\'esordonn\'es et H\'et\'erog\`enes\\
Case 86, 4 place Jussieu, F-75252 Paris Cedex 05, France
}

\affiliation{
AGH University of Science and Technology,
Faculty of Physics and Applied Computer Science\\
al. Mickiewicza 30, PL-30059 Krak\'ow, Poland
}

\date{\today}

\begin{abstract}
We study how to restore site percolation on a damaged square lattice with nearest neighbor (N$^2$) interactions. Two strategies are suggested  for a density $x$ of  destroyed sites  by a random attack at $p_c$. In the first one, a density $y$ of new sites are created with longer range interactions, either next nearest neighbor (N$^3$) or next next nearest neighbor (N$^4$).  In the second one, new longer range interactions N$^3$ or N$^4$ are created  for a fraction $v$ of the remaining $(p_c-x)$ sites in addition to their  N$^2$ interactions. In both cases, the values of $y$ and $v$ are tuned in order to restore site percolation which then occurs at new percolation thresholds, respectively $\pi_3$, $\pi_4$, $\pi_{23}$ and $\pi_{24}$. Using Monte Carlo simulations the values of the pairs $\{y, \pi_3 \}$, $\{y, \pi_4\}$ and $\{v, \pi_{23}\}$, $\{v, \pi_{24}\}$ are calculated for the whole range $0\leq x \leq p_c(\text{N}^2)$. Our schemes are applicable to all regular lattices.
\end{abstract}


\maketitle

For several decades the calculation of percolation thresholds has been an ongoing technical challenge (see \cite{recent,wier} for the most recent ones). Up to now, analytical solutions have been limited to few a very two-dimensional lattices.
However computer Monte Carlo simulations have allow to estimate percolation thresholds for an increasing large number of systems  \cite{intro,books,suding,rosowsky}.
High temperatures series expansions have been also extensively used \cite{adler}.
The most studied case is the hypercube due to its simple computer representation with available numerical estimates of  percolation threshold up to thirteen dimensions \cite{intro,sg_allpc,pcd13}.
Some universal formulas have been also suggested \cite{wier,gm}.
The percolation phenomenon  \cite{intro,books} may also occur on a large variety of complex  networks where sites are not distributed regularly \cite{networks,graphs,havlin}.
The site coordination number $z$ varies for each site with a distribution which depends mainly on the network growth rules.
For example, when subsequent sites are attached randomly to already existing ones the site degree becomes exponential \cite{graphs,exp}.
When the attachment is preferential \cite{graphs,ab} the distribution follows a power law, and for so-called classical random graphs it is given by a Poisson distribution \cite{graphs,crg}.
Such growing networks may reflect some properties of real networks --- like WWW or Internet infrastructure --- and have been the subject of  studies of their resistance on  intentional or random attacks \cite{havlin}.
At odd and for the first time we investigate  in this paper the vulnerability at $p_c$ of a regular lattice with nearest neighbor (N$^2$) interactions under a random attack. 
In particular we study how to restore site percolation on a square lattice  once a fraction $x$ of the initial $p_c$ sites have destroyed.
Two strategies are suggested.
In the first one, new sites are created with longer range interactions, either next nearest neighbor (N$^3$) or next next nearest neighbor (N$^4$). 
It is worth to note that when we consider sites with N$^3$ or N$^4$ interactions, only N$^3$ or N$^4$ interactions are active.
It means in particular that for these sites there exist no N$^2$ interactions.

\begin{figure}
\begin{center}
(a) \includegraphics[scale=.35]{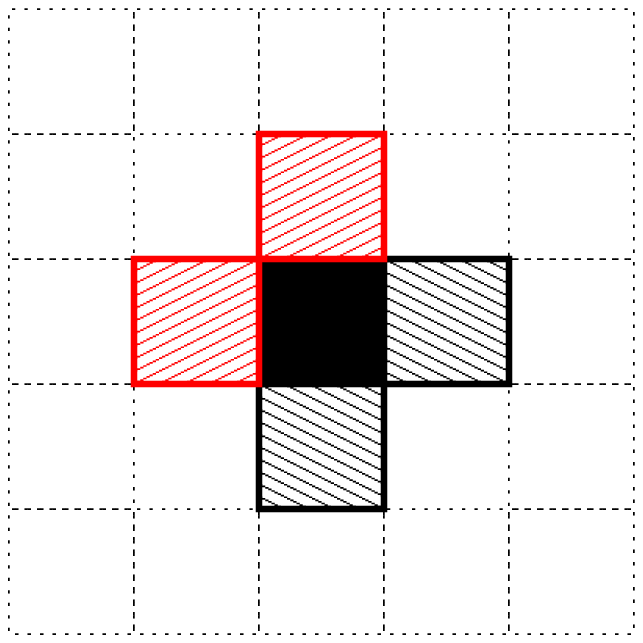}
(b) \includegraphics[scale=.35]{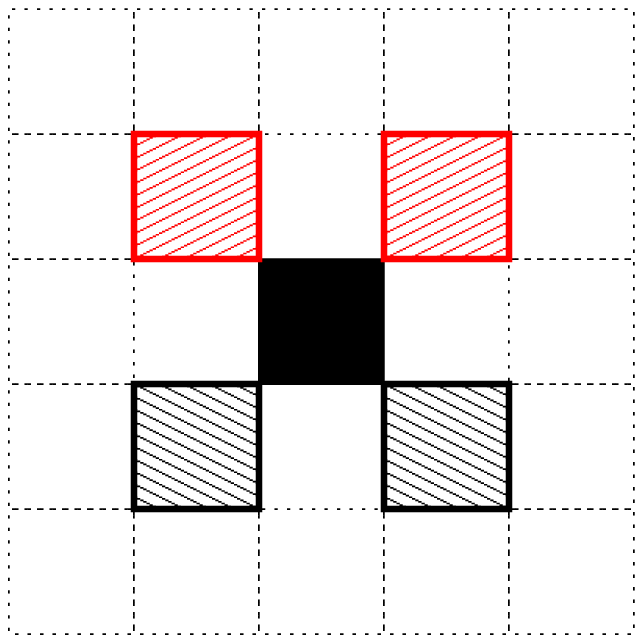}
(c) \includegraphics[scale=.35]{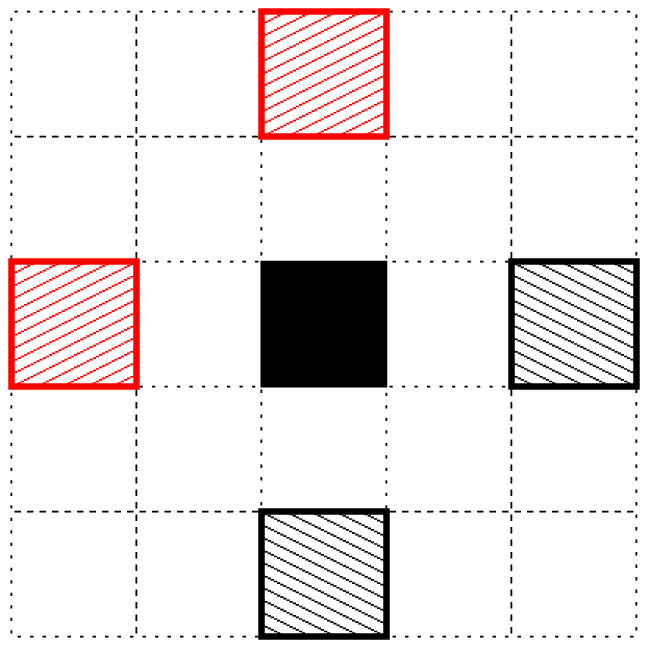}\\
(d) \includegraphics[scale=.35]{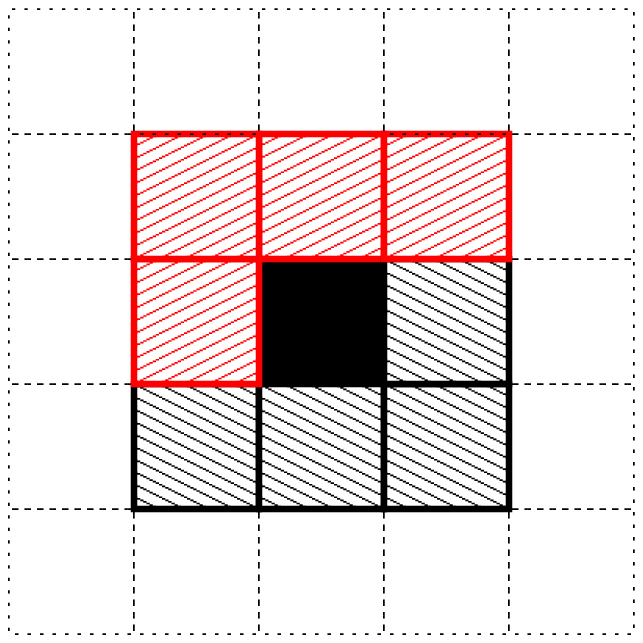}
(e) \includegraphics[scale=.35]{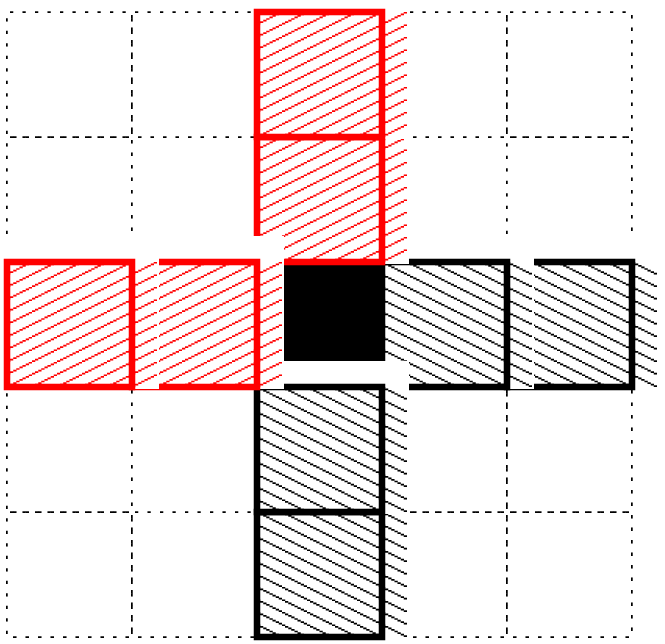}
(f) \includegraphics[scale=.35]{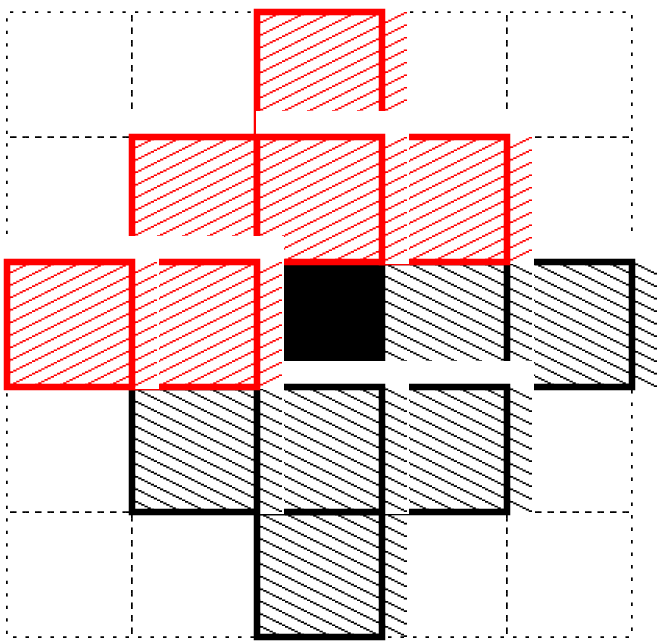}
\caption{Various sites neighborhoods on the square lattice: (a) N$^2$, (b) N$^3$,
 (c) N$^4$, and examples of their combinations: (d) N$^2$+N$^3$, (e) N$^2$+N$^4$, (f) N$^2$+N$^3$+N$^4$.}
\label{fig-ngbr}
\end{center}
\end{figure}

In the second strategy, no additional sites are created but instead new longer range interactions are created  for a fraction $v$ of the remaining not damaged after attack $(p_c-x)$ sites.  
We consider N$^3$ or N$^4$ interactions in addition to their former N$^2$ interactions.
Therefore $v(p_c-x)$ sites have N$^2$ plus either N$^3$ or N$^4$ interactions while $(1-v)(p_c-x)$ sites have only their initial N$^2$ interactions. 
Fig. \ref{fig-ngbr} shows sites with N$^2$, N$^3$, N$^4$, (N$^2$+N$^3$), (N$^2$+N$^4$) and (N$^2$+N$^3$+N$^4$) neighborhoods on the lattice.

In both cases, the values of $y$ and $v$ are tuned in order to restore site percolation which occurs then at new percolation thresholds, respectively $\pi_3$, $\pi_4$, $\pi_{23}$ and $\pi_{24}$.
Using Monte Carlo simulations the values of the pairs $\{y, \pi_3 \}$, $\{y, \pi_4\}$ and $\{v, \pi_{23}\}$, $\{v, \pi_{24}\}$ are calculated for the whole range $0\leq x \leq p_c(\text{N}^2)$.
Our both schemes are applicable to all regular lattices.

\begin{figure}
\begin{center}
(a) \includegraphics[scale=.6]{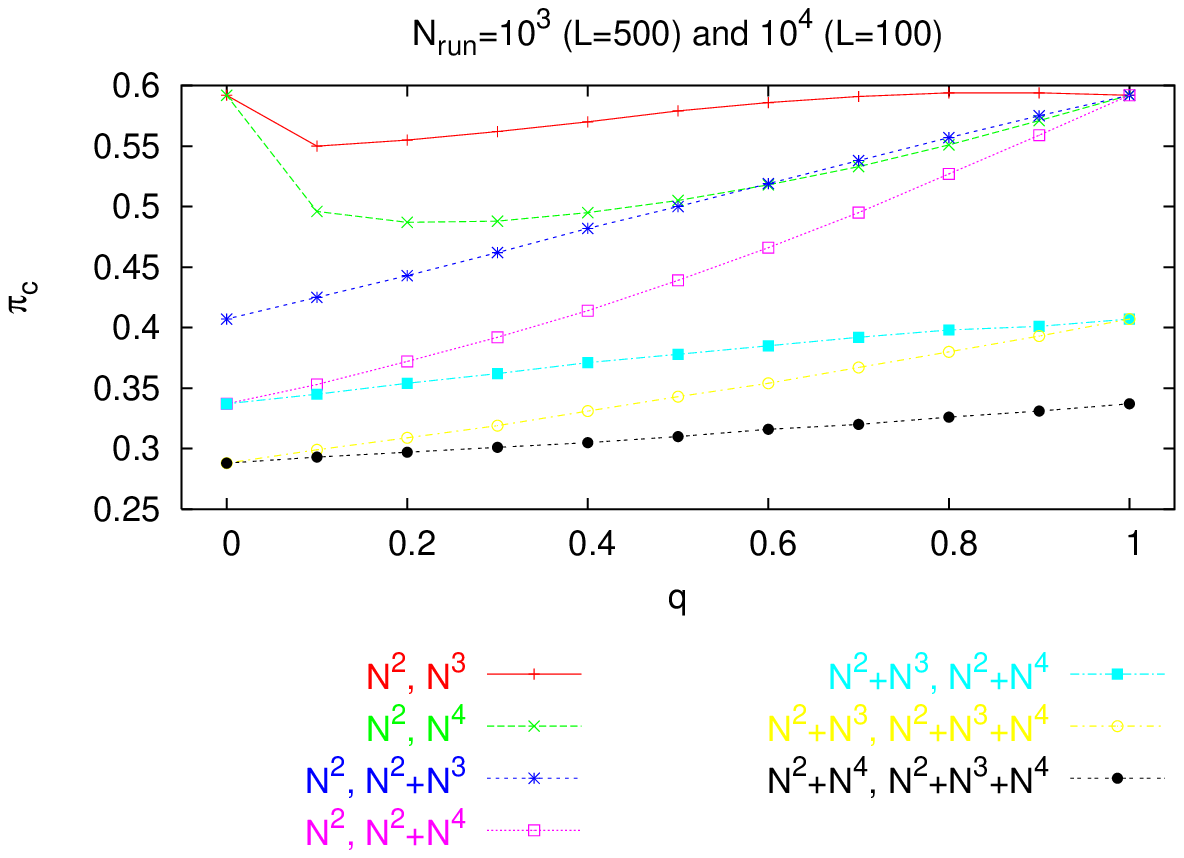}\\
(b) \includegraphics[scale=.6]{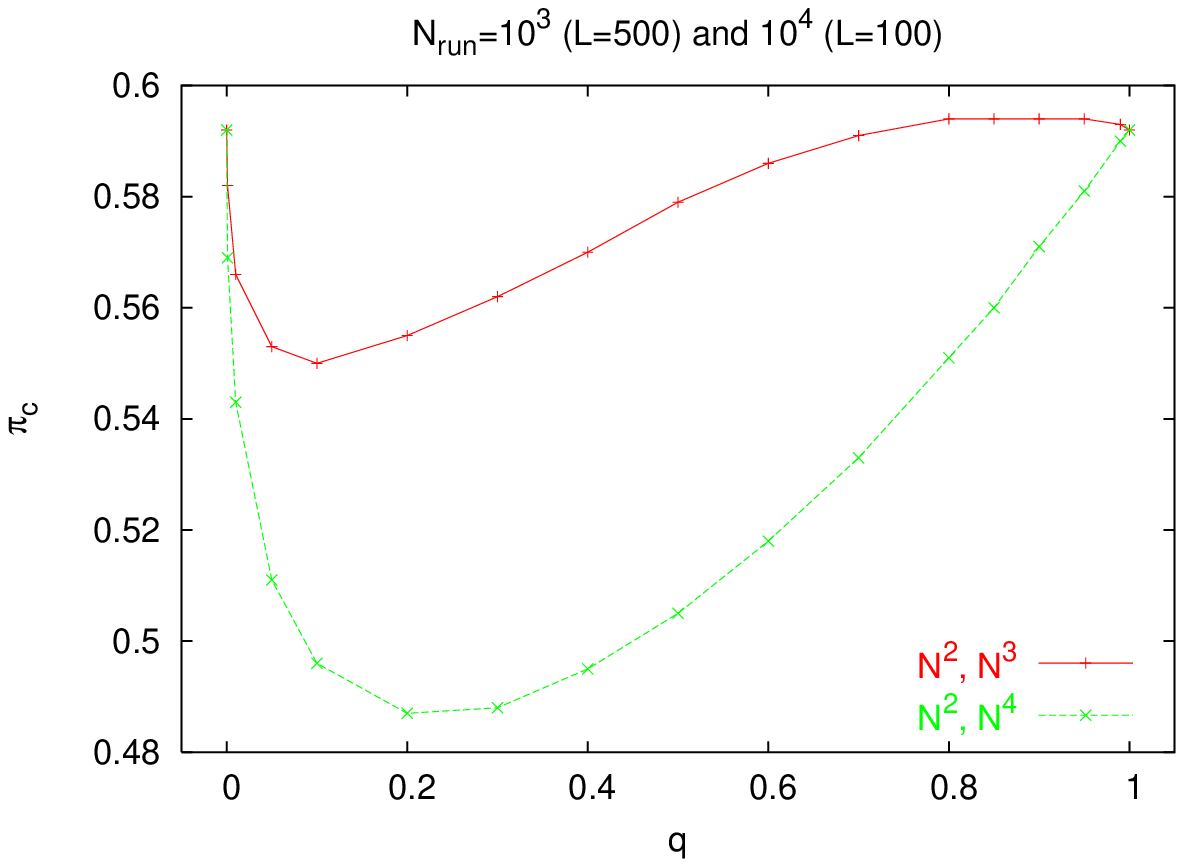}
\caption{(a) The percolation threshold $\pi_c$ dependence on the neighborhoods mixing parameter $q$.
In part (b) the results for
$\pi_3=\pi_c(q,\text{N}^2,\text{N}^3)$ and
$\pi_4=\pi_c(q,\text{N}^2,\text{N}^4)$ are repeated.
The lines are guides for eyes.}
\label{fig-q}
\end{center}
\end{figure}

However, as it  will be explained below, it is much easier, for computational reasons, to evaluate the percolation threshold on a square lattice with a given fraction $q$ of occupied sites with one kind of neighborhood while remaining fraction $(1-q)$ of occupied sites have another one.
Then, having the percolation threshold $\pi_c$ and the mixing neighborhood parameter $q$ we can easily extract thevalues of $\{x,y\}$ or $\{x,v\}$.
We also estimate percolation thresholds $\pi_c$ for the different values of the mixing parameter $q$ for various pairs of mixed neighborhoods, respectively
N$^2$ with N$^3$,
N$^2$ with N$^4$,
N$^2$ with (N$^2$+N$^3$),
N$^2$ with (N$^2$+N$^4$),
(N$^2$+N$^3$) with (N$^2$+N$^4$),
(N$^2$+N$^3$) with (N$^2$+N$^3$+N$^4$)
and
(N$^2$+N$^4$) with (N$^2$+N$^3$+N$^4$).
These thresholds are presented in Fig. \ref{fig-q}.

\bigskip

\begin{figure}
\begin{center}
\includegraphics[scale=.4]{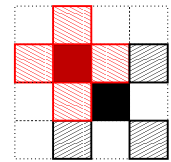}
\caption{Are full-filled black and red sites in one cluster?
The red site has N$^2$ while the black one N$^3$.
Our answer is positive, while the HKA will give such an answer only with
the probability of 50\%.
The answer is sites labeling order dependent.}
\label{fig-nn-nnn}
\end{center}
\end{figure}

To perform our calculations of percolation thresholds we are using the Hoshen--Kopelman algorithm (HKA) \cite{hka} from existing computational techniques \cite{newman,ziff,leath}. 
With HKA each occupied site gets a label.
The sites in the same cluster have the same labels and different labels are assigned to different clusters thus allowing to recognize which sites belong to which clusters.
The HKA becomes particularly efficient for the bus-bar percolation problem, i.e. when we check if the site in distance $\ell$ from the first fully occupied line is still connected to that line through the sites at the distances smaller than $\ell$.
Then a computer stores only single line of sites and may go through the lattice only once \cite{intro}.
However, when all possible links between sites are desired we have to keep the whole lattice during the simulation \cite{intro,truehka}.
The situation becomes even more complicated when we mix neighborhoods on the lattice, i.e. when different sites have different neighborhoods, as presented in Fig. \ref{fig-nn-nnn}.
There, black site \textit{is not} N$^2$ of dark (red) one, while dark (red) one \textit{is} N$^3$ of black one.
Note, that both sites \textit{are} in the same cluster.
However, the HKA will classify them as the members of one common cluster when it goes in type-writer order \footnote{from top-to-bottom and from left-to-right}, and will leave them separated when it goes in reverse-type-writer order \footnote{from bottom-to-top and from right-to-left}.
Thus, for the lattices with mixed neighborhoods we consider each site as having effectively only slashed (red) `half' of their neighborhood presented in Fig. \ref{fig-ngbr}, due to computational reasons only.

\begin{figure}
\begin{center}
\includegraphics[scale=.6]{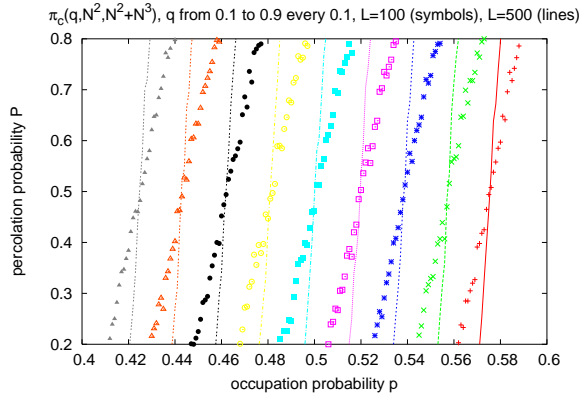}
\caption{The percolation probability $P$ dependence on the occupation probability $p$
for different values of mixing parameter $q$ which changes from 10\% to 90\% every
10\% from right to left.
The N$^2$ and (N$^2$+N$^3$) neighborhoods are mixed.
The symbols correspond to $L=100$ while lines to $L=500$.
The crossing points predict the percolation thresholds $\pi_{23}=\pi_c(q,\text{N}^2,\text{N}^2+\text{N}^3)$.}
\label{fig-pc}
\end{center}
\end{figure}

To start our simulations we denote $\pi_c(q, \text{N}_1, \text{N}_2)$ the percolation threshold for a lattice on which the fraction $q$ of occupied sites has N$_1$ neighborhood while the remaining fraction $(1-q)$ of occupied sites has N$_2$.
The percolation thresholds values $\pi_c$ are then evaluated from the crossing point of two curves showing dependences of the percolation probability $P$ vs. the sites occupation probability $p$ for lattices of linear sizes $L=100$ and $500$ (see Fig. \ref{fig-pc}).
The results are averaged over $N_{\text{run}}=10^3$ and $10^4$ for $L=500$ and $100$, respectively. They are presented in Fig. \ref{fig-q}.
As a matter of fact, each site has only `half' of their neighborhood as we have explained earlier.

\bigskip

\begin{figure}
\begin{center}
(a) \includegraphics[scale=.6]{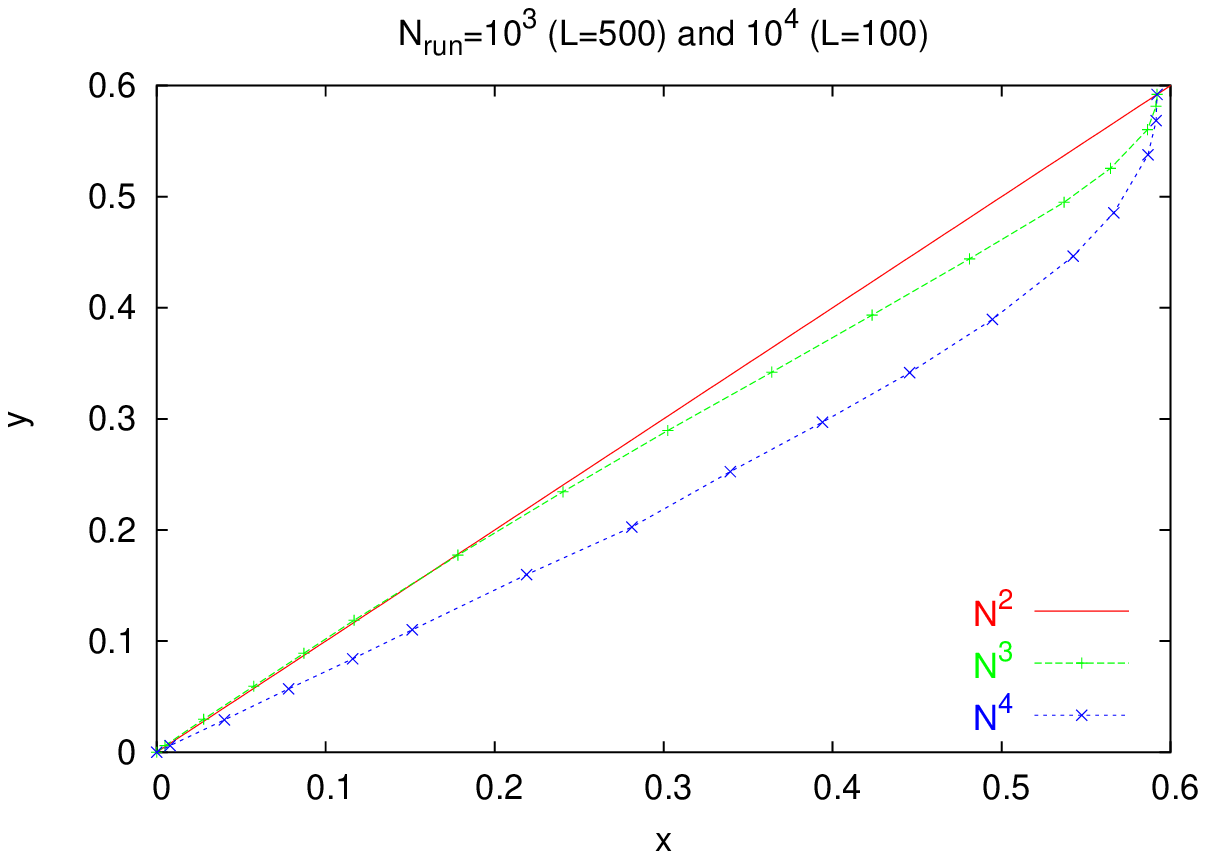}\\
(b) \includegraphics[scale=.6]{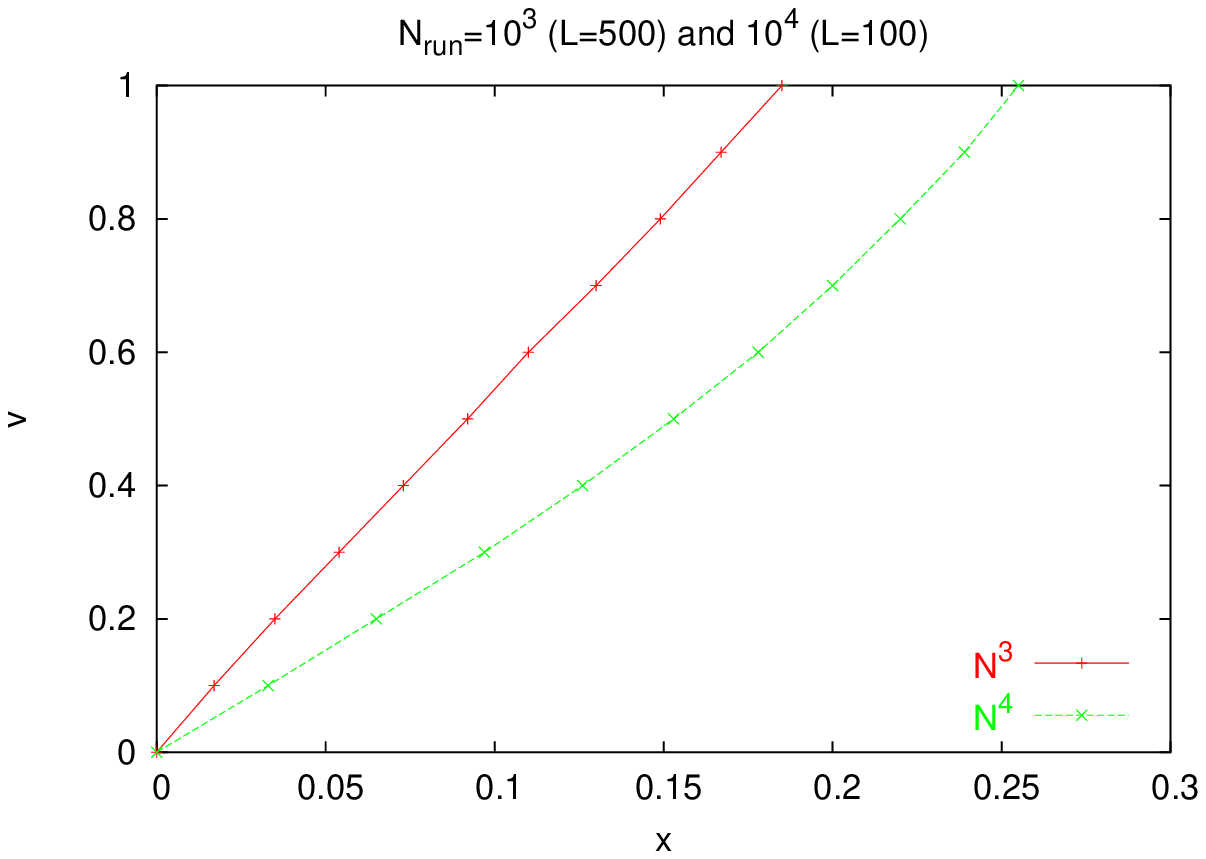}
\caption{The fraction of sites which must be (a) reoccupied ($y$)
or (b) enriched with long-range interactions ($v$)
to recover percolation phenomenon when the fraction $x$ of sites with the N$^2$ was emptied.
The lines are guides for eyes.}
\label{fig-del-add}
\end{center}
\end{figure}

At this stage we are  in position to determine which site density $y$ with long-range interactions (for instance N$^3$ or N$^4$) is necessary to reconstruct a percolation phenomenon which occurs now at $\pi_3\equiv\pi_c(q,\text{N}^2,\text{N}^3)$ or $\pi_4\equiv\pi_c(q,\text{N}^2,\text{N}^4)$, when the fraction $x$ of the occupied sites with the N$^2$ neighborhoods has been removed due to the random attack.
The results for selected values of parameter $q$ are presented in Fig. \ref{fig-q}(b).
They may be easily transformed to answer that question for both N$^3$ and N$^4$ additional interactions.
To do that, we estimate the percolation threshold $\pi_c$ for given pair of neighborhoods (here N$^2$ with N$^3$ and N$^2$ with N$^4$) and given ratio $q/(1-q)$ of their homogeneous mixture on the lattice.
For estimated value of $\pi_c$ we find immediately the fraction $y=(1-q)\pi_c$ of sites with long-range interactions.
The remaining fraction of occupied sites, which is $q\pi_c=p_c(\text{N}^2)-x$, has N$^2$ kind of neighborhood.
The results are collected in Tabs. \ref{tab-nnn} and \ref{tab-nnnn} and presented in Fig. \ref{fig-del-add}(a).
\begin{table}
\caption{The fraction $y$ of sites with N$^3$ which must be add to reconstruct the percolation phenomenon --- which occurs at $\pi_3=\pi_c(q,\text{N}^2,\text{N}^3)$ --- when only $[p_c(\text{N}^2)-x]$ of sites with N$^2$ is occupied.}
\label{tab-nnn}
\begin{ruledtabular}
\begin{tabular}{llll}
$1-q$   & $x$   & $y$   & $\pi_3$ \\
\hline
0.0     & 0.0000        & 0.0000        & 0.592 \\
0.1     & 0.0574        & 0.0594        & 0.594 \\
0.3     & 0.1783        & 0.1773        & 0.591 \\
0.4     & 0.2404        & 0.2344        & 0.586 \\
0.5     & 0.3025        & 0.2895        & 0.579 \\
0.6     & 0.3640        & 0.3420        & 0.570 \\
0.8     & 0.4810        & 0.4440        & 0.555 \\
0.95    & 0.5645        & 0.5254        & 0.553 \\
0.99    & 0.5863        & 0.5603        & 0.566 \\
1.0     & 0.5920        & 0.5920        & 0.592 \\
\end{tabular}
\end{ruledtabular}
\end{table}
\begin{table}
\caption{The fraction $y$ of sites with N$^4$ which must be add to reconstruct the percolation phenomenon --- which occurs at $\pi_4=\pi_c(q,\text{N}^2,\text{N}^4)$ --- when only $[p_c(\text{N}^2)-x]$ of sites with N$^2$ is occupied.}
\label{tab-nnnn}
\begin{ruledtabular}
\begin{tabular}{llll}
$1-q$   & $x$           & $y$           &$\pi_4$\\
\hline
0.0     & 0.0000        & 0.0000        & 0.592 \\
0.05    & 0.0401        & 0.0291        & 0.581 \\
0.2     & 0.1512        & 0.1102        & 0.551 \\
0.3     & 0.2189        & 0.1599        & 0.533 \\
0.4     & 0.2812        & 0.2072        & 0.518 \\
0.5     & 0.3395        & 0.2525        & 0.505 \\
0.7     & 0.4456        & 0.3416        & 0.488 \\
0.9     & 0.5424        & 0.4464        & 0.496 \\
0.99    & 0.5866        & 0.5376        & 0.543 \\
1.0     & 0.5920        & 0.5920        & 0.592 \\
\end{tabular}
\end{ruledtabular}
\end{table}

The other strategy which may be applied for the reconstruction process is to enrich the fraction $v$ of sites of those which survived an attack by additional long-range bonds to other existing and not damaged sites.
Again, numerically we deal with the square lattice on which a given fraction $q$ of occupied sites has N$^2$ neighborhoods, while remaining $(1-q)$ has N$^2$+N$^3$ or N$^2$+N$^4$.
The latter number is also the fraction $v$ of sites with long interaction among all occupied sites.
If the percolation for such mixture occurs at $\pi_c$ then all of these sites have N$^2$ in their neighborhoods. 
The fraction $x$ which is missed to the $p_c(\text{N}^2)$ is then simply $p_c(\text{N}^2)-\pi_c$.
The results are collected in Tab. \ref{tab-bonds} and presented in Fig. \ref{fig-del-add}(b).
\begin{table}
\caption{The fraction $v$ of sites with N$^2$ which must be enriched with N$^3$ or N$^4$ bonds to reconstruct the percolation phenomenon --- which occurs at respectively $\pi_{23}$ or $\pi_{24}$ --- when only $[p_c(\text{N}^2)-x]$ of sites with N$^2$ was saved after attack.}
\label{tab-bonds}
\begin{ruledtabular}
\begin{tabular}{llllll}
$1-q$ & $x_{23}$ & $\pi_{23}$ & $x_{24}$ & $\pi_{24}$ & $v$ \\
\hline
0.0 & 0.000 & 0.592 & 0.000 & 0.592 & 0.0 \\
0.1 & 0.017 & 0.575 & 0.033 & 0.559 & 0.1 \\
0.2 & 0.035 & 0.557 & 0.065 & 0.527 & 0.2 \\
0.3 & 0.054 & 0.538 & 0.097 & 0.495 & 0.3 \\
0.4 & 0.073 & 0.519 & 0.126 & 0.466 & 0.4 \\
0.5 & 0.092 & 0.500 & 0.153 & 0.439 & 0.5 \\
0.6 & 0.110 & 0.482 & 0.178 & 0.414 & 0.6 \\
0.7 & 0.130 & 0.462 & 0.200 & 0.392 & 0.7 \\
0.8 & 0.149 & 0.443 & 0.220 & 0.372 & 0.8 \\
0.9 & 0.167 & 0.425 & 0.239 & 0.353 & 0.9 \\
1.0 & 0.185 & 0.407 & 0.255 & 0.337 & 1.0 \\
\end{tabular}
\end{ruledtabular}
\end{table}

For `pure' cases, when each site has the same neighborhoods (i.e. when
$q=0$ or $q=1$) the results agree very well with thresholds $p_c$ recently presented in Ref. \onlinecite{kmsg}.
Similar checks for bi-colored lattices with N$^2$ and N$^3$ neighborhoods were performed in Ref. \onlinecite{bulg}. 
For sites reconstruction process, when $x=0$ or $x=p_c(\text{N}^2)$, the difference between originally used N$^2$ sites and their N$^3$ and/or N$^4$ substitutes vanishes:
\begin{itemize}
\item In the limit $x\to 0$ the damages are very rare and thus reconstruction process is not too costly independently what kind of `bricks' (sites with N$^3$ or N$^4$) is used.
\item On the other hand, when $x\to p_c(\text{N}^2)$ the system must be completely reconstructed starting from `zero'.
Since all of `pure' N$^2$, N$^3$ and N$^4$ lattices has the same percolation threshold
$p_c(\text{N}^2)= p_c(\text{N}^3)= p_c(\text{N}^4)= 0.5927460$
\cite{pcNN,kmsg} the cost of reaching the percolation threshold becomes again neighborhoods independent.
\item Note, that in both cases (N$^3$ and N$^4$) there is an optimal ratio $q/(1-q)$ for which the percolation threshold $\pi_c$ is the lowest.
\item For N$^3$ sites used for the reconstruction process $\pi_3 > p_c(\text{N}^2)$ when $0.8<q<1$, i.e. when damages are relatively small ($x<0.2$).
In that case addition sites with diagonal bonds do not much help with restoring the percolation but increase total sites density.
\end{itemize}

For the second scenario, there is a critical density of sites $x_c$ above which damages are not compensable by adding long range bonds even to all of remaining sites.
For $x>x_c$ the sites concentration falls below the threshold $p_c$ for the square lattice with all sites having the same mixed neighborhood, for instance N$^2$+N$^3$ or N$^2$+N$^4$ \cite{kmsg}.
The critical values $x_c(\text{N})=\pi_c(0,\text{N}^2,\text{N}^2+\text{N})-p_c(\text{N}^2)$ are 0.185, 0.255 and 0.304 for N=N$^3$, N$^4$ and N$^3$+N$^4$, respectively.

For both reconstruction strategies, the percolation occurs more easily, it is achieved at a lower cost, if neighborhood with larger radius are employed as presented in Fig. \ref{fig-del-add}.

\bigskip

To conclude, in contrast to Ref. \cite{havlin} we have not been interested on how to destroy connectivity on a given lattice but rather on how to effectively reconstruct it after a random destruction.
Using Monte Carlo simulations we have reported square lattice site percolation thresholds $\pi_c$ for given neighborhoods characterized by a mixing parameter $q$ and various pairs of mixed neighborhoods build from basics ones, i.e. N$^2$, N$^3$ and N$^4$.
We have showed quantitatively that `repairing costs' are lower when sites with longer radius of neighborhood are employed for site reconstruction process at intermediate range of damages.
The strategy involving bond enrichment fails if  damages are too large.
The critical damages size $x_c$  depends on the percolation threshold $p_c(\text{N}^2+\text{N})$, where N is the kind of bonds used in the enriching process.
Both our schemes are readily applicable to all regular lattices. The results may found useful applications in the fields of  econophysics  \cite{econo} and sociophysics \cite{socio}, for instance to study reconstruction processes in social systems after an exposion to some sort of damages where interpersonal bonds have been broken.

\begin{acknowledgments}
K.M.'s stay at the Universit\'e Pierre et Marie Curie et CNRS was financed by the European Science Foundation with grant No. COST-\-STSM-\-P10-\-00262.
The numerical calculations were carried out in ACK\---CY\-F\-RO\-NET\---AGH.
The machine time on SGI 2800 is financed by the Polish Ministry of Science and Information Technology under grant No. MNiI/\-SGI2800/\-AGH/\-049/\-2003.
\end{acknowledgments}


\end{document}